# THE IGEX $^{76}$Ge NEUTRINOLESS DOUBLE-BETA DECAY EXPERIMENT: PROSPECTS FOR NEXT GENERATION EXPERIMENTS

C. E. Aalseth,[1,†] F. T. Avignone III,[1] R. L. Brodzinski,[2] S. Cebrian,[3] E. Garcia,[3] D. Gonzalez,[3] W. K. Hensley,[2] I. G. Irastorza,[3] I. V. Kirpichnikov,[4] A. A. Klimenko,[5] H. S. Miley,[2] A. Morales,[3] J. Morales,[3] A. Ortiz de Solorzano,[3] S. B. Osetrov,[5,‡] V. S. Pogosov,[6] J. Puimedon,[3] J. H. Reeves,[2] M. L. Sarsa,[3] A. A. Smolnikov,[5] A. S. Starostin,[4] A. G. Tamanyan,[6] A. A. Vasenko,[4] S. I. Vasiliev,[5] and J. A. Villar[3]

[1]*University of South Carolina, Columbia, SC 29208 USA*

[2]*Pacific Northwest National Laboratory, Richland, WA 99352 USA*

[3]*University of Zaragoza, 50009 Zaragoza, Spain*

[4]*Institute for Theoretical and Experimental Physics, 117259 Moscow, Russia*

[5]*Institute for Nuclear Research, Baksan Neutrino Observatory, 361609 Neutrino, Russia*

[6]*Yerevan Physical Institute, 375036 Yerevan, Armenia*

(The IGEX Collaboration)

---

[†] Present address: Pacific Northwest National Laboratory, Richland, WA 99352 USA.
[‡] Present address: BioTraces Inc, Herndon, VA 20171 USA.

**Abstract**

The International Germanium Experiment (IGEX) has analyzed 117 mole yr of $^{76}$Ge data from its isotopically enriched (86% $^{76}$Ge) germanium detectors. Applying pulse-shape discrimination (PSD) to the more recent data, the lower bound on the half-life for neutrinoless double-beta decay of $^{76}$Ge is: $T_{1/2}(0\nu) > 1.57 \times 10^{25}$ yr (90% C.L.). This corresponds to an upper bound in the Majorana neutrino mass parameter, $\langle m_\nu \rangle$, between 0.33 eV and 1.35 eV, depending on the choice of theoretical nuclear matrix elements used in the analysis.

**INTRODUCTION**

Neutrino oscillation experiments have produced "smoking guns" for non-zero neutrino mass in the solar neutrino deficit[1], in the excess of $p(\bar{\nu}_e, e^+)n$ reactions from the LSND experiment[2], and more recently from the strong zenith-angle dependence of the electron to muon event ratio in the SuperKamiokande (SK) data[3] (see also [4-6]). The results of reactor neutrino experiments[7] constrain the disappearance of $\bar{\nu}_e$ well enough to imply that the SK data are dominated by $\nu_\mu \to \nu_\tau$ ($\bar{\nu}_\mu \to \bar{\nu}_\tau$) oscillations, with only a minimal oscillation to electron-type neutrinos, since reactor experiments show that they do not oscillate as readily as required by the SK data.

While the interpretation of the SK data in terms of neutrino oscillations is widely accepted, there have been many questions concerning the interpretation of the LSND data as evidence of $\bar{\nu}_\mu \to \bar{\nu}_e$ oscillations, as well as doubts that the standard solar model was accurate enough to support the conclusion that there is really a deficit of solar neutrinos. When the results of all solar neutrino experiments are considered together, there is no scenario in which these data are compatible with the standard solar model unless the flux of $\nu_e$ from the sun oscillates partially into other ν-flavors to which the experiments are not sensitive.

On 17 July 2001, however, the Sudbury Neutrino Observatory (SNO) collaboration settled this issue. They reported their results from the direct measurement of the reaction rate of $d(\nu_e, e^-)pp$ from solar neutrinos[8]. The solar neutrino flux implied from these data was compared with that implied from the neutral-current component of the neutrino-electron elastic scattering data from Super K. It was concluded that there is an active non-electron flavor neutrino component in the solar neutrino flux, and that the total flux of active neutrinos from the $^8$B reaction is in close agreement with the standard solar model of Bahcall and his coworkers[9]. The standard solar model is thereby confirmed, and the case for neutrino oscillations is now compelling.

The final question is that of the LSND positive indication of $\bar{\nu}_\mu \to \bar{\nu}_e$ oscillations. All attempts to incorporate these results in the same analysis with those from the solar neutrino and atmospheric neutrino experiments fail in the context of any scenario involving only three neutrino flavors.

Accepting as fact that now both solar and atmospheric neutrino experiments give clear evidence for neutrino oscillations, there are only two conclusions that can be drawn from the LSND data, assuming the interpretation of these data is accurate. Either the excess events from the reaction $p(\bar{\nu}_e, e^+)n$ in the LSND are due to phenomena other than $\bar{\nu}_\mu \to \bar{\nu}_e$ oscillation, or there must exist a fourth generation of neutrinos. This generation must be "sterile" with respect to "normal" weak interactions[10]. To insist on accepting one or the other of these options at the present time is to accept an unsubstantiated theoretical prejudice. This issue is still very much an open one. In any case, it is safe to conclude at this point that neutrinos do possess properties outside of the standard model of particle physics.

While an unambiguous interpretation of all of the above neutrino oscillation experiments is not yet possible, it is abundantly clear that neutrinos exhibit mass and flavor mixing. Accordingly, sensitive searches for neutrinoless double-beta $(0\nu\beta\beta)$ decay are more important than ever. Experiments with kilogram quantities of germanium, isotopically enriched in $^{76}$Ge, have thus far proven to be the most sensitive, specifically the Heidelberg-Moscow[11] and IGEX[12] experiments. The resulting half-life lower limits, $1.9 \times 10^{25}$ y [11] and $1.6 \times 10^{25}$ y [12], imply that a new generation of experiments will be required to make significant improvements in sensitivity, as will be discussed later.

According to the standard solar model of Bahcall and co-workers [9], the deficit in the solar-$\nu_e$ flux on earth can be explained by the Mikheyev-Smirnov-Wolfenstein (MSW) resonant oscillation. Until recently, the acceptable regions in the parameter space, $\delta m^2 - \sin^2 2\theta$, were believed to be incompatible with neutrino masses that would allow direct observation of 0νββ decay. Petcov and Smirnov[13] have shown that both MSW and vacuum oscillation solutions of the solar neutrino problem can be compatible with 0νββ decay driven by an effective Majorana electron-neutrino mass in the range 0.1 to 1.0 eV. The interpretation of all the neutrino oscillation data together, as discussed later, implies a range that could be between 5 and 10 times lower. The exploration of such a range will require next-generation experiments. Some that are being proposed are: CAMEO[14], CUORE[15], EXO[16], GENIUS[17], Majorana[18], and MOON[19]. Very brief descriptions of each of these are given later.

In this article the results of the analysis of 117 fiducial mole yr of $^{76}$Ge data from the IGEX experiment are discussed, and an attempt is made to project the requirements of next generation 0νββ–decay experiments to advance the state of the art in sensitivity well beyond the two presently most sensitive experiments[11,12].

## DOUBLE-BETA DECAY

Neutrinoless double-beta decay is the only known way to determine if neutrinos are Majorana particles. According to Kayser, et al.[20], the observation of 0νββ decay would constitute unambiguous proof that at least one neutrino eigenstate has non-zero mass, when interpreted in the context of any gauge theory. Some insight into this issue can also be obtained from the black-box theorem of Schechter and Valle[21]. There are many comprehensive reviews of double-beta decay in the literature[22].

The decay rate for this process involving the exchange of a Majorana neutrino can be written as follows:

$$\lambda_{\beta\beta}^{0\nu} = G^{0\nu}(E_0, Z)\langle m_\nu \rangle^2 \left| M_f^{0\nu} - (g_A/g_V)^2 M_{GT}^{0\nu} \right|^2 . \qquad (1)$$

In equation (1), $G^{0\nu}$ is the two-body phase-space factor including coupling constants; $M_f^{0\nu}$ and $M_{GT}^{0\nu}$ are the Fermi and Gamow-Teller nuclear matrix elements, respectively; and $g_A$ and $g_V$ are the axial-vector and vector relative weak coupling constants, respectively. The quantity $\langle m_\nu \rangle$ is the effective Majorana neutrino mass given by:

$$\langle m_\nu \rangle \equiv \left| \sum_{k=1}^{2n} \lambda_k^{CP} (U_{ek}^L)^2 m_k \right| , \qquad (2)$$

where $\lambda_k^{CP}$ is the CP eigenvalue associated with the k[th] neutrino mass eigenstate (±1 for CP conservation); $U_{ek}^L$ is the (e,k) matrix element of the transformation between flavor eigenstates $|\nu_\ell\rangle$ and mass eigenstates $|\nu_k\rangle$ for left-handed neutrinos;

$$|\nu_\ell\rangle = \sum U_{\ell k}|\nu_k\rangle ,\qquad(3)$$

and $m_k$ is the mass of the k$^{th}$ neutrino mass eigenstate. A Feynman diagram of the process is shown in Figure 1.

Neutrinoless double-beta decay has been hypothesized as driven by a number of other mechanisms: intrinsic right-handed currents, the emission of Goldstone bosons (Majorons), and the exchange of supersymmetric particles; however, here only the process involving Majorana neutrino mass will be discussed.

The effective Majorana neutrino mass, $\langle m_\nu \rangle$, is directly derivable from the measured half-life of the decay as follows:

$$\langle m_\nu \rangle = m_e \left(F_N T_{1/2}^{0\nu}\right)^{-1/2} \text{ eV} ,\qquad(4)$$

where $F_N \equiv G^{0\nu} \left| M_f^{0\nu} - (g_A/g_V)^2 M_{GT}^{0\nu} \right|^2$, and $m_e$ is the electron mass. This quantity derives from nuclear structure calculations and is model dependent as seen in Table 1.

## NUCLEAR STRUCTURE CONSIDERATIONS

From eq. (1), the sensitivity of a given experiment to the parameter, $\langle m_\nu \rangle$, depends directly on nuclear matrix elements. In this regard, $2\nu\beta\beta$-decay experiments have some value in testing models, although $M^{2\nu}$ and $M^{0\nu}$ are completely different. The weak-coupling shell-model calculations of Haxton, et al.[23], were an extensive effort to explain the geochemical double-beta decay half-lives of $^{128}$Te, $^{130}$Te, and $^{82}$Se, as well as to predict the half-life of $^{76}$Ge. These early calculations used the value $(g_A/g_V) = 1.24$. It was later realized that a value of unity is more appropriate for a neutron decaying in a complex nucleus. The shell-model prediction then became $T_{1/2}^{2\nu}\left(^{82}Se\right) = 0.8 \times 10^{20}$ yr, which is within 20% of the TPC value measured by the Irvine group[24].

In 1986, the CalTech group[25] introduced the Quasiparticle Random Phase Approximation (QRPA) with three parameters to account for pairing, particle-hole, and particle-particle interactions. Later, similar models were developed by the Tübingen group[26] and the Hiedelberg group[27]. In all of these models, the parameter, $g_{pp}$, characterizing the short-range particle-particle correlations, had a single value near which the 2νββ-decay matrix elements vanish. However, it is generally agreed that the 0νββ-decay matrix elements have a much softer dependence on these parameters and thus are more stable. In 1994, Faessler concluded that the inclusion of neutron-proton pairing interactions reduces the dependence of $M_{GT}^{2\nu}$ on $g_{pp}$ [28]. More recently, however, new large-space shell-model calculations by Caurier, et al.[29], yielded significantly different results, as shown in Table 1. This is an important open question yet to be understood.

## THE IGEX EXPERIMENT

A complete description of the IGEX experiment has been published[30] with results from analyzing ~75 mole yr of $^{76}$Ge data. An additional 41.9 mole yr have been added; the totals are presented in Table 2 and Figure 2. The darkened spectrum in Fig. 2 results from applying PSD to about 15% of the 75 mole yr data set and to the entire 41.9 mole yr data set. Detailed models of the crystal and associated first stage preamplifier have been constructed, and pulse shapes from various sources of background were simulated. The PSD analysis leading to the results shown in Fig. 2 is a very conservative visual technique that compared experimental pulse shapes to computed single-site and multi-site pulses.

Using standard statistical techniques, there are fewer than 3.1 candidate events (90% C.L.) under a peak having FWHM = 4 keV and centered at 2038.56 keV. This corresponds to:

$$T_{1/2}^{0\nu}\left(^{76}\text{Ge}\right) > \frac{4.87 \times 10^{25} \text{ yr}}{3.1} \cong 1.6 \times 10^{25} \text{ yr}. \tag{4}$$

The values of $F_N$ given in Table 1 lead to $0.3 \leq \langle m_\nu \rangle \leq 1.3\,\text{eV}$. Readers can interpret the data given in Table 2 as they wish.

The continuation of the on-going $^{76}$Ge experiments can improve these bounds; however, very probably not into the regime of $\langle m_\nu \rangle$ implied by the present neutrino oscillation data interpreted with current conventional wisdom. The authors' views of what future 0νββ-decay experiments would have to achieve, based on a current understanding of all of the neutrino experiments, are discussed below.

**REQUIREMENTS OF FUTURE 0νββ-DECAY EXPERIMENTS BASED ON PROBABLE NEUTRINO SCENARIOS**

The SuperKamiokande data imply maximal mixing of $\nu_\mu$ with $\nu_\tau$ and $\delta m_{23}^2 \cong (55 \text{ meV})^2$. The solar neutrino data from SK also imply that the small mixing angle solution to the solar neutrino problem is disfavored, and that $\delta m^2 \text{(solar)} \cong (7 \text{ meV})^2$. Based on these interpretations, one probable scenario for the neutrino mixing matrix has, at least approximately, the following form[31]:

$$\begin{pmatrix} \nu_e \\ \nu_\mu \\ \nu_\tau \end{pmatrix} = \begin{pmatrix} 1/\sqrt{2} & 1/\sqrt{2} & 0 \\ -1/2 & 1/2 & 1/\sqrt{2} \\ 1/2 & -1/2 & 1/\sqrt{2} \end{pmatrix} \begin{pmatrix} \nu_1 \\ \nu_2 \\ \nu_3 \end{pmatrix}. \tag{5}$$

The neutrino masses can be arranged in two hierarchical patterns in which $\delta m_{31}^2 \cong \delta m_{32}^2 \sim (55\,meV)^2$ and $\delta m_{21}^2 \sim (7\,meV)^2$. With the available data, it is not possible to determine which hierarchy, $m_3 > m_1(m_2)$ or $m_1(m_2) > m_3$, is the correct one, nor the absolute value of any of the mass eigenstates. The two possible schemes are depicted in Fig. 3.

The consideration of reactor neutrino and atmospheric neutrino data together strongly indicates that the atmospheric neutrino oscillations are dominantly $\nu_\mu \to \nu_\tau \,(\overline{\nu}_\mu \to \overline{\nu}_\tau)$, which implies, as seen from eq. (5), that $\nu_e$ is a mixture of $\nu_1$ and $\nu_2$. In the chosen case, where $U_{e3} = 0$, eq. (2) only contains one relative CP phase, $\varepsilon$, and reduces to:

$$\langle m_\nu \rangle = \tfrac{1}{2}(m_1 + \varepsilon m_2), \qquad (6)$$

whereas the large mixing angle solution of the solar neutrino problem implies

$$(m_2^2 - m_1^2) = (7\,meV)^2. \qquad (7)$$

Consideration of bi-maximal mixing yields four cases to be analyzed: (a) $m_1 \cong 0$, (b) $m_1 \gg 7$ meV, (c) $m_3 \cong 0$ and (d) the existence of a mass scale, M, where M >> 55 meV.

a) If $m_1 = 0$, $m_2 \cong 7\,meV$, and $\langle m_\nu \rangle = \dfrac{m_2}{2}$.

b) If $m_1 \gg 7\,meV \equiv M$ and $\langle m_\nu \rangle \cong \dfrac{M}{2}(1+\varepsilon)$.

c) If $m_3 = 0$, $m_1 \cong m_2 \cong 55\,meV$, and $\langle m_\nu \rangle \cong 0$ or $55\,meV$.

d) If $M \gg 55\,meV$, $m_1 \cong m_2 \cong (M + 55\,meV)$, and $\langle m_\nu \rangle \cong \dfrac{m_1}{2}(1+\varepsilon)$.

If we assume that $\varepsilon \cong +1$, and that neutrinos are Majorana particles, then it is very probable that $\langle m_\nu \rangle$ lies between 7 meV and the present bound from $^{76}$Ge 0νββ-decay experiments.

The requirements for a next generation experiment can easily be deduced by reference to eq. (8):

$$T_{1/2}^{0\nu} = \frac{(\ln 2)Nt}{c}, \qquad (8)$$

where $N$ is the number of parent nuclei, $t$ is the counting time, and $c$ is the upper limit on the number of $0\nu\beta\beta$-decay counts consistent with the observed background. To improve the sensitivity of $\langle m_\nu \rangle$ by a factor of 100, the quantity $Nt/c$ must be increased by a factor of $10^4$. The quantity $N$ can feasibly be increased by a factor of $\sim 10^2$ over present experiments, so that $t/c$ must also be improved by that amount. Since practical counting times can only be increased by a factor of 2 to 4, the background should be reduced by a factor of 25 to 50 below present levels. These are approximately the target parameters of the next generation neutrinoless double-beta decay experiments.

Georgi and Glashow give further motivation for these increased-sensitivity, next-generation double-beta decay experiments[32]. They discuss six "facts" deduced from atmospheric neutrino experiments [3-5] and from solar neutrino experiments [1], with constraints imposed by reactor experiments [7]. From these they conclude that if neutrinos play an essential role in the large-scale structure of the universe, the six facts "are mutually consistent if and only if solar neutrino oscillations are nearly maximal." They further state that stronger bounds on neutrinoless double-beta decay could constrain solar neutrino data to only allow the just-so oscillations.

## NEXT GENERATION EXPERIMENTS

The CAMEO proposal involves placing isotopically enriched parent isotopes at the center of BOREXINO. One example given involves 65 kg of $^{116}$CdWO$_4$ scintillation crystals. The collaboration predicts a sensitivity of $\langle m_\nu \rangle \sim 60$ meV, and with 1000 kg the prediction is $\langle m_\nu \rangle \sim 20$ meV [14].

CUORE is a proposed cryogenic experiment with 25 towers of 40 detectors, each a 750-g $TeO_2$ bolometer. This detector would utilize natural abundance Te, containing 33.8% $^{130}$Te. A pilot experiment, CUORICINO, comprising one CUORE tower, is under construction. With equivalent background, CUORE would be as sensitive as 400 to 950 kg of Ge enriched to 86% $^{76}$Ge, depending on the nuclear matrix elements used to derive $\langle m_\nu \rangle$. It will be performed in Gran Sasso[15].

EXO is a large proposed TPC, either high-pressure gas or liquid, of enriched $^{136}$Xe. This novel technique involves schemes for locating, isolating, and identifying the daughter $^{136}$Ba$^+$ ion by laser resonance spectroscopy. A program of research and development is underway at the Stanford linear accelerator[16].

GENIUS is a proposal to use between 1.0 and 10 tons of "naked" germanium detectors, isotopically enriched to 86% in $^{76}$Ge, directly submerged in a large tank of liquid nitrogen functioning both as a cooling method and a clean shield. Extensive studies were made based on certain assumptions, measurements, and Monte-Carlo simulations. In ref. [17], the authors claim a sensitivity range of 1-10 meV for $\langle m_\nu \rangle$, using $10^3$-$10^4$ kg of enriched Ge. A research and development program is underway in the Gran Sasso Laboratory to develop the techniques for cooling and operating "naked" Ge detectors in liquid nitrogen for extended periods[17].

The Majorana Project is a proposed significant expansion of the IGEX experiment, utilizing newly-developed segmented detectors along with pulse-shape discrimination (PSD) techniques that have been developed since the data presented in this paper were obtained. It proposes 500 fiducial kg of Ge isotopically enriched to 86% in $^{76}$Ge in the form of 200-250 detectors. Each detector will be segmented into 12 electrically-independent volumes, each of which will be instrumented with the new PSD system. A prototype is near completion and will be installed underground in 2002.

The MOON experiment is a proposed major extension of the ELEGANTS experiment. It will utilize between 1 and 3 tons of Mo foils, isotopically enriched to 85% in $^{100}$Mo, inserted between plastic scintillators. It will have coincidence and tracking capabilities to search for 0νββ decay as well as solar neutrinos. This novel technique for detecting solar neutrinos depends on the special properties of the nuclear decay schemes of $^{100}$Mo and its daughters, allowing both event and background identification[19].

This list of proposals should produce several experiments with the sensitivity to actually observe 0νββ decay, or obtain upper bounds on $\langle m_\nu \rangle$ reaching the sensitivity range implied by recent neutrino oscillation results. The IGEX[12] and Heidelberg-Moscow[11] $^{76}$Ge experiments not only yield the best current bounds on $\langle m_\nu \rangle$, they also provide most of the technology needed in future $^{76}$Ge experiments.

In the above discussions of the range of $\langle m_\nu \rangle$ that could render neutrinoless double-beta decay observable, one scenario was chosen out of a number of possibilities. There have been several extensive discussions of various other interpretations of neutrino oscillation data and their impact on the range of probable values of this important parameter[33-38].

In one case[38], it was found that for three-neutrino mixing, $|\langle m \rangle| \sim 10$ meV if the neutrino mass spectrum is hierarchical. On the other hand, if two of the neutrino eigenstates are quasi-degenerate, with $m_1$ having a small mass, $|\langle m \rangle|$ could be as large as 100 meV. In this case, early stages of next generation experiments could directly observe neutrinoless double-beta decay.

Reference [37] predicts ranges of 1-1000 meV for the Majorana mass parameter considering all possible solar neutrino solutions, including the cases of hierarchy, partial degeneracy, and inverse hierarchy.

References [33] through [37] discuss the impact on CP-violation in the neutrino sector and its connection to neutrino oscillations, tritium beta-decay, and double-beta decay experiments. Reference [34] discusses three- and four-neutrino flavor scenarios in the context of next-generation tritium beta-decay measurements and double-beta decay experiments. It also discusses how these data could help determine the pattern of neutrino mass eigenstates, and possibly the relative CP-violating phase, in the case that two neutrino states are involved in solar neutrino oscillations.

Frequently appearing publications on the subject almost always refer to the importance of conducting next-generation neutrinoless double-beta decay experiments. A complete understanding of the neutrino mass matrix depends on three types of data: neutrino oscillations, tritium beta decay measurements, and neutrinoless double-beta decay. Each is analogous to one leg of a three-legged stool, and each is necessary for the complete picture. The case for a significant investment in next-generation experiments of all three types is compelling.

**ACKNOWLEDGMENTS**

The authors are grateful to Petr Vogel for his views on the probable neutrino scenarios used as examples in this text. The Canfranc Astroparticle Underground Laboratory is operated by the University of Zaragoza under Contract AEN96-1657. This research was partially funded by the Spanish Commission for Science and Technology (CICYT), the U.S. National Science

[38]  S. M. Bilenky, C. Guinti, W. Grimus, B. Kayser, and S. T. Petcov, arXiv: hep – Ph / 9907234 v3, 30 August 1999; Phys. Lett. **B 465**, 193 (1999).

**Tables**

## Table 1

Nuclear structure factors $F_N$ and Majorana neutrino mass parameters $\langle m_\nu \rangle$ for a $0\nu\beta\beta$ decay half-life of $1.57 \times 10^{25}$ years.

| $F_N$ (years$^{-1}$) | Model | $\langle m_\nu \rangle$ (eV) |
|---|---|---|
| $1.56 \times 10^{-13}$ | Shell model [23] | 0.33 |
| $9.67 \times 10^{-15}$ | QRPA [25] | 1.35 |
| $1.21 \times 10^{-13}$ | QRPA [26] | 0.38 |
| $1.12 \times 10^{-13}$ | QRPA [27] | 0.38 |
| $1.41 \times 10^{-14}$ | Shell model [29] | 1.09 |

## Table 2

IGEX $^{76}$Ge data after the partial application of PSD for 117 mole yr. The starting energy of each 2-keV bin is given.

| Energy | Events | Energy | Events |
|---|---|---|---|
| 2020 | 2.9 | 2042 | 5.5 |
| 2022 | 9.1 | 2044 | 6.0 |
| 2024 | 3.4 | 2046 | 1.7 |
| 2026 | 2.0 | 2048 | 5.3 |
| 2028 | 4.6 | 2050 | 3.4 |
| 2030 | 6.5 | 2052 | 4.6 |
| 2032 | 2.3 | 2054 | 5.0 |
| 2034 | 0.6 | 2056 | 0.6 |
| 2036 | 0.0 | 2058 | 0.1 |
| 2038 | 2.0 | 2060 | 4.3 |
| 2040 | 1.5 | | |

**Figures**

# Figure 1
Feynman diagram of 0νββ decay process.

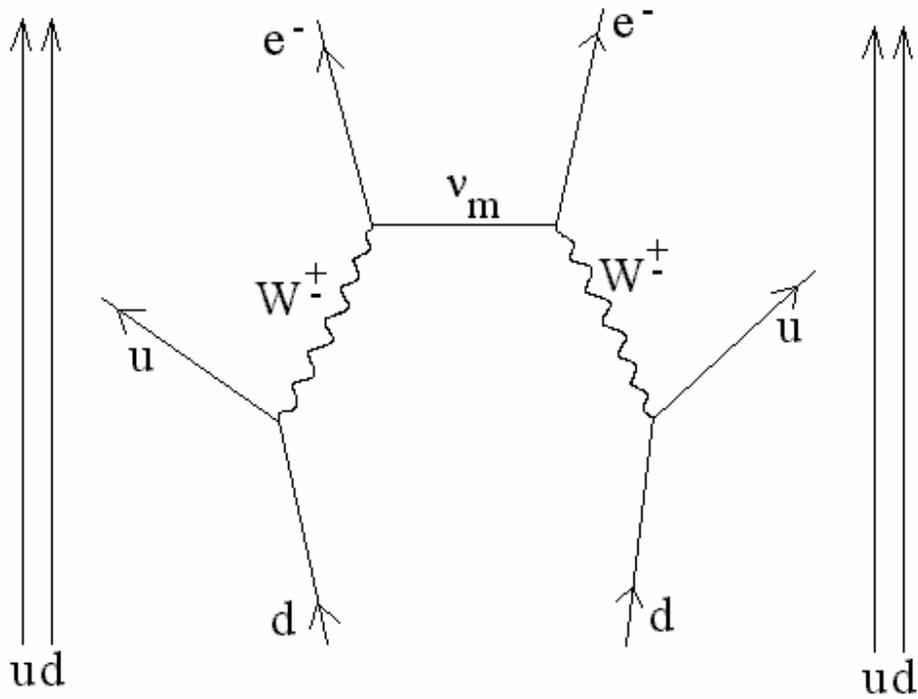

**Figure 2**

Histograms from 117 effective mole yr of IGEX $^{76}$Ge data. Energy bins are labeled on the left edge. The darkened spectrum results from the application of PSD to ~45% of the total data set. The gaussian curve represents the 90% CL constraint of ≤ 3.1 0νββ-decay events and has a FWHM of ~4 keV, corresponding to the effective energy resolution of the entire experiment.

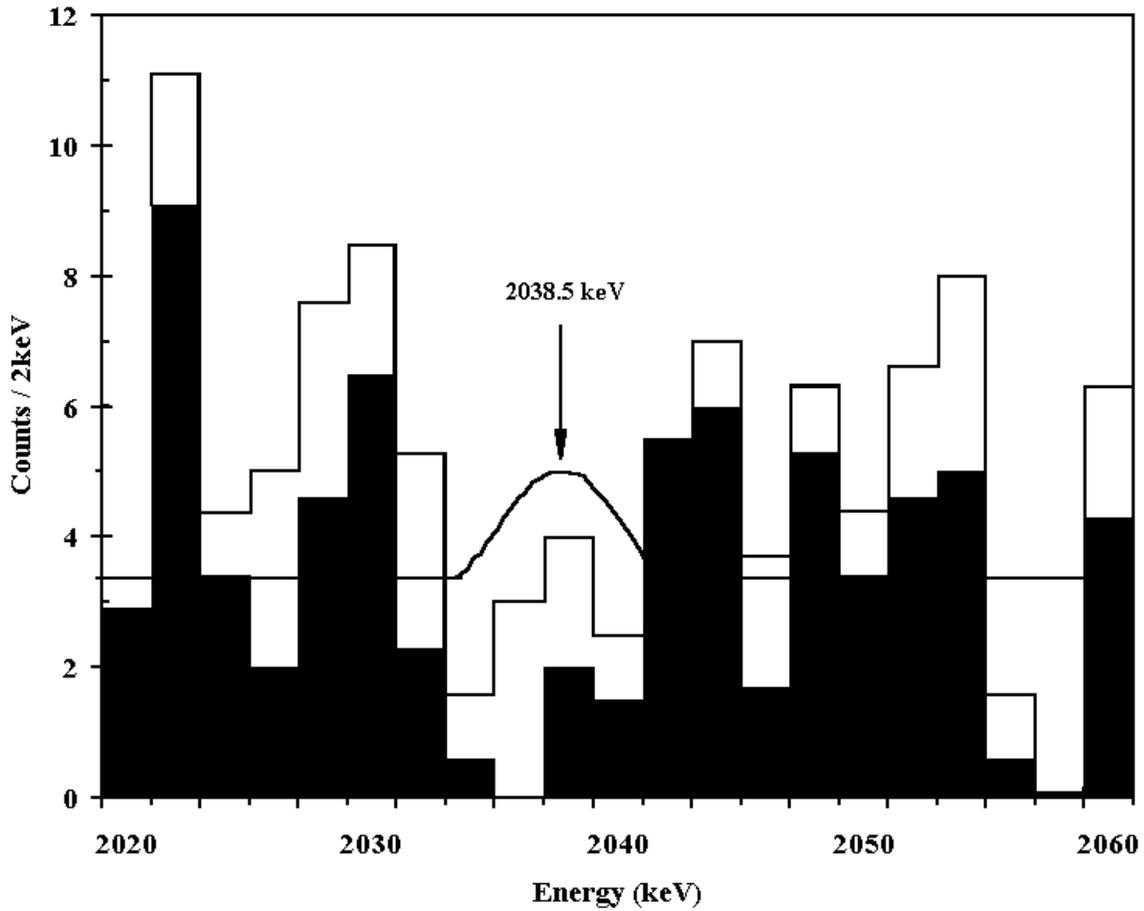

**Figure 3**
Two possible neutrino mass hierarchies.

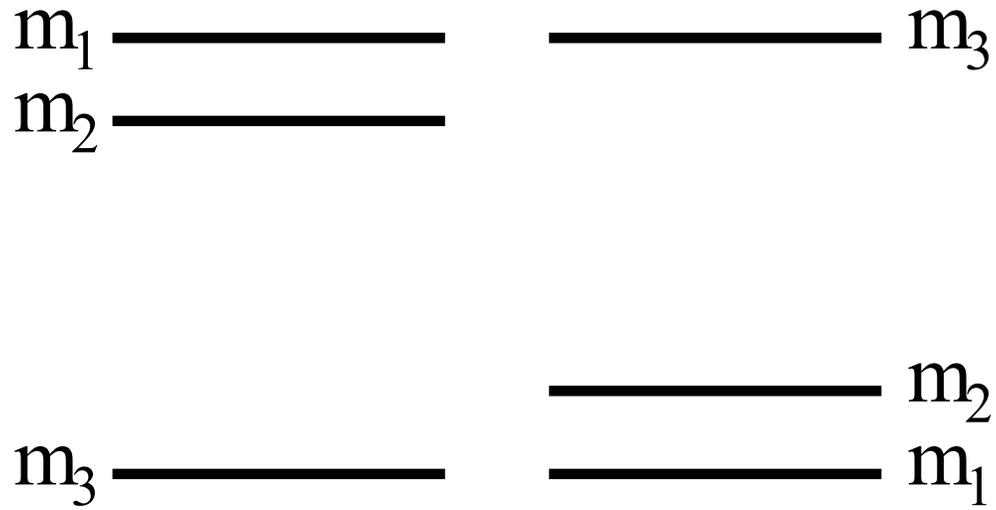